# A large population of galaxies 9 to 12 billion years back in the history of the Universe

O. Le Fèvre[1], S. Paltani[1], S. Arnouts[1], S. Charlot[2,3], S. Foucaud[4], O. Ilbert[1,5], H. J. McCracken[3,12], G. Zamorani[6], D. Bottini[4], B. Garilli[4], V. Le Brun[1], D. Maccagni[4], J. P. Picat[7], R. Scaramella[8], M. Scodeggio[4], L. Tresse[1], G. Vettolani[8], A. Zanichelli[8], C. Adami[1], S. Bardelli[6], M. Bolzonella[5], A. Cappi[6], P. Ciliegi[6], T. Contini[7], P. Franzetti[4], I. Gavignaud[7,9], L. Guzzo[10], A. Iovino[10], B. Marano[5], C. Marinoni[10], A. Mazure[1], B. Meneux[1], R. Merighi[6], R. Pellò[7], A. Pollo[10], L. Pozzetti[6], M. Radovich[11], E. Zucca[6], M. Arnaboldi[11], M. Bondi[8], A. Bongiorno[5], G. Busarello[11], L. Gregorini[8], F. Lamareille[7], G. Mathez[7], Y. Mellier[3,12], P. Merluzzi[5], V. Ripepi[11] & D. Rizzo[7]

[1]*Laboratoire d'Astrophysique de Marseille, OAMP, UMR6110, CNRS–Université de Provence Aix-Marseille I, BP8, F-13376 Marseille Cedex 12, France.*
[2]*Max Planck Institut fur Astrophysik, D-85741 Garching, Germany.*
[3]*Institut d'Astrophysique de Paris, UMR 7095, 98 bis Bvd Arago, F-75014 Paris, France.*
[4]*IASF-INAF - via Bassini 15, I-20133, Milano, Italy.*
[5]*Università di Bologna, Dipartimento di Astronomia - Via Ranzani 1, I-40127, Bologna, Italy.*
[6]*INAF-Osservatorio Astronomico di Bologna - Via Ranzani 1, I-40127, Bologna, Italy.*
[7]*Laboratoire d'Astrophysique de l'Observatoire Midi-Pyrénées (UMR 5572) - 14, avenue E. Belin, F-31400 Toulouse, France.*
[8]*IRA-INAF - Via Gobetti 101, I-40129, Bologna, Italy.*
[9]*European Southern Observatory, Karl-Schwarzschild-Strasse 2, D-85748 Garching bei Munchen, Germany.*
[10]*INAF-Osservatorio Astronomico di Brera - Via Brera 28, Milan, Italy.*
[11]*INAF-Osservatorio Astronomico di Capodimonte - Via Moiariello 16, I-80131, Napoli, Italy.*
[12]*Observatoire de Paris, LERMA, 61 Avenue de l'Observatoire, F-75014 Paris, France.*

**To understand the evolution of galaxies, we need to know as accurately as possible how many galaxies were present in the Universe at different epochs[1]. Galaxies in the young Universe have hitherto mainly been identified using their expected optical colours[2–4], but this leaves open the possibility that a significant population remains undetected because their colours are the result of a complex mix of stars, gas, dust or active galactic nuclei. Here we report the results of a flux-limited I-band survey of galaxies at look-back times of 9 to 12 billion years. We find 970 galaxies with spectroscopic redshifts between 1.4 and 5. This population is 1.6 to 6.2 times larger than previous estimates[2–4], with the difference increasing towards brighter magnitudes. Strong ultraviolet continua (in the rest frame of the galaxies) indicate vigorous star formation rates of more than 10–100 solar masses per year. As a consequence, the cosmic star formation rate is higher than previously measured at redshifts of 3 to 4.**

One technique to identify galaxies in the distant Universe is to use the discontinuity in the spectrum produced by the photon absorption below the 912 Å Lyman limit to isolate high-redshift galaxies from the dominating population of foreground galaxies in colour–colour diagrams. This technique has led to the identification of the Lyman-break galaxies (LBGs) at redshifts $z \approx 2$–4 (refs 2–4), the current reference for the high-redshift population. However, this colour-selection technique with its associated uncertainties and assumptions could exclude some significant fraction of the high-redshift galaxy population. The question therefore naturally arises whether there are galaxies at similar distances that have escaped searches at optical wavelengths. This can be checked using the basic observational approach of selecting all galaxies in a region of the sky brighter than a given apparent flux limit, followed by a systematic spectroscopic redshift measurement. This magnitude-selection technique has been used successfully from $z \approx 0$ up to $z \approx 1.5$ (refs 5–9), but assembling significant samples of a few hundred $z > 2$ galaxies requires spectroscopic observations of several thousands of sources.

Here we report the spectroscopic identification of galaxies with $1.4 \leq z \leq 5$ from a purely I-band (~8,100 Å) flux-selected sample. The VIMOS VLT (Very Large Telescope) Deep Survey (VVDS) has assembled a 'first-epoch' sample of ~8,000 galaxies in the VVDS field in the Cetus constellation with measured redshifts in the range $0 \leq z \leq 5$ (ref. 10). There are 970 galaxies with $1.4 \leq z \leq 5$; the spectra clearly show all the features that are expected in this rest-frame wavelength range (Fig. 1). Using the ultraviolet continuum flux at 1,500 Å rest wavelength ($L_{1500}$, in erg s$^{-1}$ Hz$^{-1}$), and the conversion[11] to star formation rate (SFR, in solar masses per year, $M_\odot$ yr$^{-1}$) given by SFR=$1.4 \times 10^{-28} L_{1500}$, we find that these galaxies have SFR=$(10$–$100) M_\odot$ yr$^{-1}$, without correction for dust absorption, indicating a strong star formation activity. Interestingly, only a small fraction (~15%) of these high-redshift spectra show Lyα in emission, with an average Lyα equivalent width of only 3 Å, a property already observed in other samples[12].

We estimate from these data the number density of galaxies in the redshift range $1.4 \leq z \leq 5$ as the number of galaxies corrected for the fraction of galaxies we have observed in the field, ~27%, and for the estimated fraction of galaxies with incorrect redshifts (Fig. 1), divided by the observed area, which is 1,700 arcmin$^2$. The galaxy surface density is reported in Table 1 for each redshift range, and presented in Fig. 2. An additional uncertainty in these estimates is due to the large-scale fluctuations in the distribution of galaxies. We checked the amplitude of these variations in the field by computing surface densities in sub-fields of sizes from 850 down to 100 arcmin$^2$. We find differences of up to a factor of 3 in the smaller fields, but only about 3% in the larger fields; these are smaller than the quoted statistical errors of our density measurements and consistent with simulation predictions[13]. Importantly, the number density values reported here can be regarded as conservative lower limits, as the galaxies that have the right colours to be LBG candidates[3], but for which we could not assign a redshift, were discarded. If all of these objects are at the redshift suggested by their colours, they could increase the projected galaxy density by additional factors of 1.5 to 2 at the fainter magnitudes at $z\approx3$ and $z\approx4$, respectively. In addition, we do not correct our density estimates for those galaxies which, while being in the redshift ranges considered here, may have been wrongly classified at lower redshift. With this conservative approach, we find a total surface density of galaxies down to a magnitude $I_{AB}$=24 of $\Sigma(z=[1.4,2])=0.758\pm0.101$, $\Sigma(z=[2.7,3.4])=0.235\pm0.025$ and $\Sigma(z=[3.7,4.5])=0.063\pm0.020$ galaxies arcmin$^{-2}$, using the same redshift ranges as in previous studies for comparison[3,4]. The galaxy surface densities we measure in the VVDS at $z\approx3$ are between 1.6 and 6.2 times larger than those obtained from the colour-selected Lyman-break samples[3].

This result therefore shows that the Universe contained more galaxies at redshifts 1.4 to 5 than previously reported using the colour-selection technique. The probability that the VVDS galaxy surface densities are consistent with the previously reported surface density of LBGs is less than $10^{-4}$ for the whole sample. Remarkably, we find that in the R-band magnitude intervals $22.5\leq R\leq23.0$, $23.0\leq R\leq23.5$ and $23.5\leq R\leq24.0$, the VVDS galaxy surface density is higher than that of colour-selected LBGs by factors of 6.2, 2.1 and 1.6 respectively, with corresponding probabilities of only 0.3%, 0.2% and 4.3% that the surface densities from the two samples are consistent with each other. This indicates that the bright end of the galaxy luminosity function at these early epochs is more populated than previously thought. The VVDS galaxies in the ($u-g$,$g-r$) diagram are mainly located near the boundary used to isolate LBGs in the colour-selection technique, as shown in Fig. 3. A significant fraction of galaxies has relatively red $g-r$ colours, which may indicate the contribution of a reddened active galactic nucleus (AGN) to this population. Steidel et al.[3] claim that the fraction of $z\approx3$ galaxies that erroneously escape the LBG selection at magnitude $I_{AB}$=24 is about 40%, a figure highly dependent on the assumptions about the properties of the high-redshift population. Our results demonstrate that magnitude-selected samples, although requiring a large number of redshift measurements, provide a more complete census of the high-redshift galaxy population. More details about the properties of this new population of bright high-redshift galaxies (BHZGs) will be given elsewhere (S.P. et al., manuscript in preparation).

While the VVDS provides direct evidence for higher galaxy densities at large redshifts than previously thought, this may have already been indirectly hinted at. From photometric-redshift analysis, Pascarelle et al.[14] reported finding more galaxies outside the LBG colour box than expected, although their result is weakened by their large photometric error bars; Foucaud et al.[15] reported a large surface density of LBGs from colour–colour selection, which is difficult to interpret as due to contamination by galaxies at other redshifts. An interesting result is the report of very bright LBGs at $z\approx3$ discovered in the Sloan Digital Sky Survey[16]: although this sample is small and might be partially contaminated by AGN, our extrapolated counts to bright magnitudes are consistent with the number density of the SDSS objects of $1.2\times10^{-6}$ galaxies arcmin$^{-2}$ at magnitude $I_{AB}$=20. Other searches of LBGs or Lyα emitters have produced samples of faint galaxies in fields about 20 times smaller, with little overlap with our luminosity range[17], making a comparison with our sample difficult. The galaxies we have identified may only be partly related to dim red galaxies[18] as they have a surface density 6 to 20 times smaller than the VVDS galaxies, and are dimmer and redder than galaxies in our survey. We also estimate from redshift surveys of the faint sub-millimetre population that sub-millimetre galaxies are about 10 times less numerous[19], and are therefore mainly a different population. We conclude that we have observed a population of bright galaxies that has not been identified before.

Our discovery of a significantly larger galaxy population at high redshift than was previously believed has important consequences for our understanding of galaxy evolution. The galaxy formation process has to spawn a more numerous galaxy population producing more stars than previously assumed at a time when the Universe was ~10–20% of its current age. Summing the ultraviolet luminosity at 1,700 Å of galaxies brighter than a rest-frame absolute magnitude $M_{AB}(1,700\text{ Å})=-22$ in the observed volume (without correction for interstellar dust absorption), we find an ultraviolet luminosity density LD(1700Å)=$1.6\times10^{25}$ erg s$^{-1}$ Hz$^{-1}$ Mpc$^{-3}$ and $1.4\times10^{25}$ erg s$^{-1}$ Hz$^{-1}$ Mpc$^{-3}$ at $z\approx3$ and $z\approx4$, respectively (for the concordance cosmology $\Omega$=0.3, $\Lambda$=0.7 and $H_0$=70 km s$^{-1}$ Mpc$^{-1}$). This is more than twice that quoted in previous studies[3,20,21]. The total number of stars formed in bright galaxies at this epoch may therefore be two to three times higher than previously inferred.

**Acknowledgements** We thank ESO for continuous support of this programme; the CNRS, the University of Provence and the Italian INAF for funding; and S.J. Lilly and A. Renzini for discussions. The observations reported here are based on observations obtained at the European Southern Observatory Very Large Telescope, and on data products produced at TERAPIX and the Canadian Astronomy Data Centre as part of the Canada-France-Hawaii Telescope Legacy Survey, a collaborative project of NRC and CNRS.

**Author Information** Reprints and permissions information is available at npg.nature.com/reprintsandpermissions. The authors declare no competing financial interests. Correspondence and requests for materials should be addressed to O.L.F. (Olivier.LeFevre@oamp.fr).

# Table 1: Projected galaxy density in the VVDS

| $I_{AB}$ magnitude | $z=[1.4,2]$ | $z=[2.7,3.4]$ | $z=[3.7,4.5]$ |
|---|---|---|---|
| 22.0–22.5 | 0.021±0.006 | 0.005±0.002 | 0.001±0.001 |
| 22.5–23.0 | 0.084±0.013 | 0.027±0.005 | 0.005±0.003 |
| 23.0–23.5 | 0.227±0.033 | 0.060±0.008 | 0.013±0.005 |
| 23.5–24.0 | 0.433±0.061 | 0.147±0.016 | 0.049±0.016 |

The number of galaxies per square arc minute in 0.5 magnitude intervals is tabulated for 3 redshift intervals from $z=1.4$ to $z=4.5$. The galaxy densities have been corrected for the sampling rate of the VVDS and the estimated fraction of incorrect redshifts.

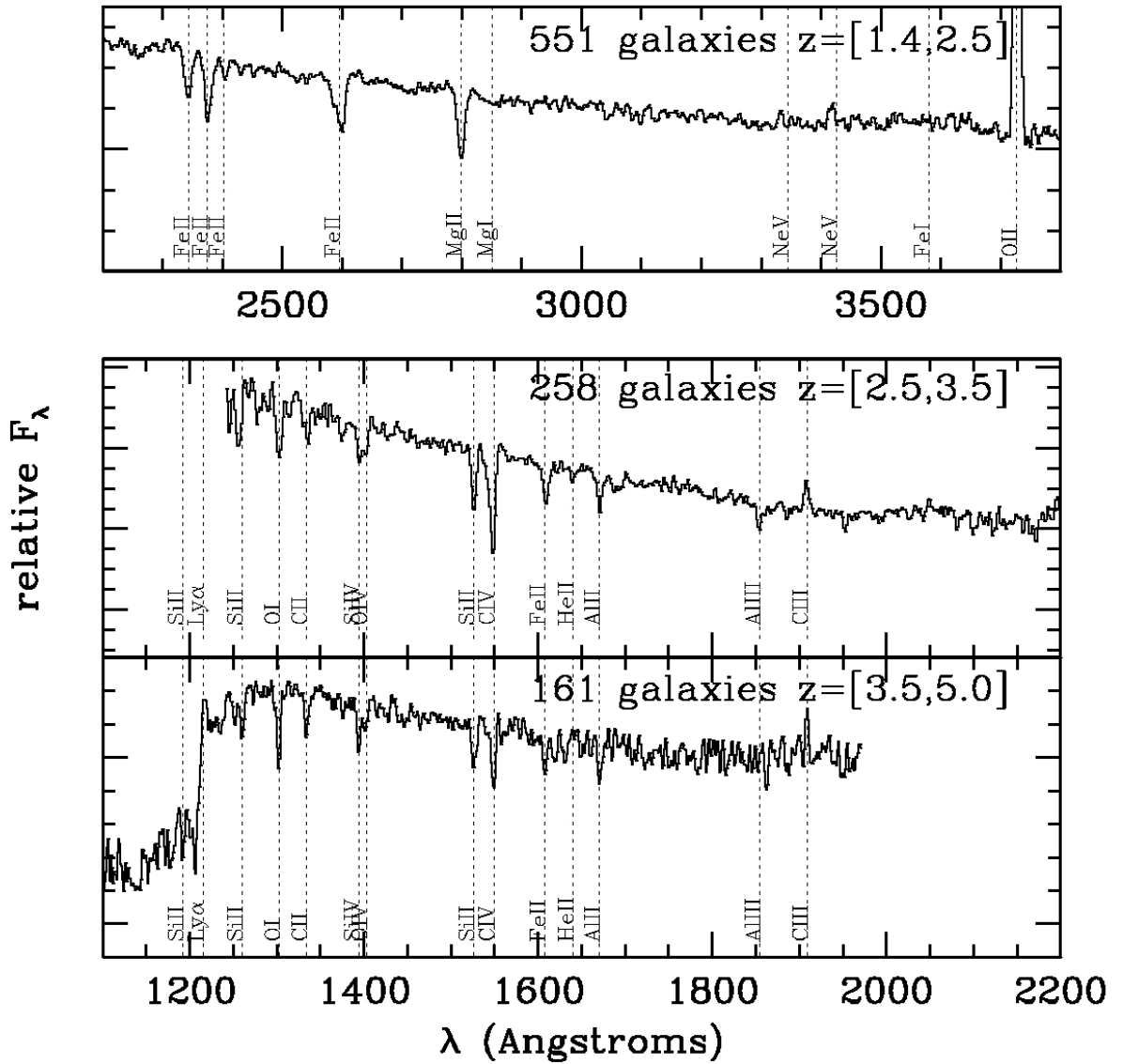

**Figure 1. Average spectra of VVDS galaxies**. The average spectra of VVDS galaxies in three redshift ranges (**a**, $1.4 \leq z < 2.5$; **b**, $2.5 \leq z < 3.5$; and **c**, $3.5 \leq z < 5.0$). The main spectral features expected at these wavelengths are clearly visible in all spectra. There are 532 galaxies with reliable redshift measurements (quality flags 2, 3 and 4; ref. 10), and 438 of lower quality (flag 1; ref. 10). Since there are a number of uncertain redshift determinations in the VVDS, we must estimate the contamination of our sample by galaxies outside the redshift range of interest, which have been erroneously included in this range during our redshift measurement process. From the comparison of the sum of stellar absorption equivalent widths in the stacked spectra of galaxies with low-quality redshift measurements to those in the stacked spectra of galaxies with secure redshifts[10], we estimate that approximately 69%, 52% and 46% of all VVDS redshifts in the range $z$=[1.4,2], [2.7,3.4] and [3.7,4.5], respectively, must be correct. $F_\lambda$, flux at wavelength $\lambda$.

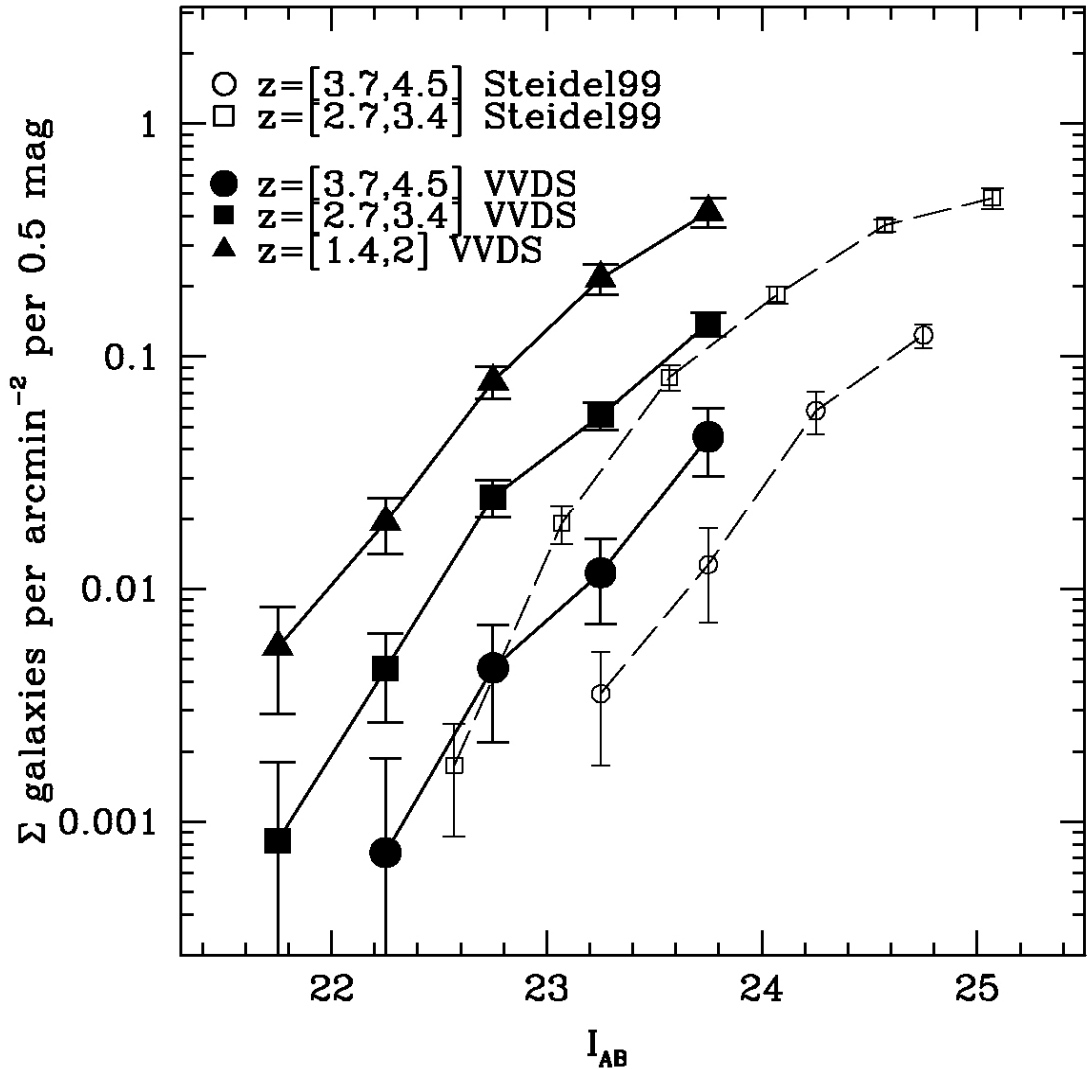

**Figure 2. Galaxy surface density.** The VVDS galaxy surface density is computed in 3 redshift domains, $z$=[1.4,2] (filled triangles), $z$=[2.7,3.4] (filled squares) and $z$=[3.7,4.5] (filled circles), compared to the surface density measured by Steidel *et al.*[2-4] (open symbols and dashed lines). The VVDS surface density is consistently larger by factors 1.6–6.2 at $z\approx3$, and by factors 2–3.5 at $z\approx4$. The surface densities plotted in this figure have been corrected for the survey sampling rate and the estimated fraction of incorrect redshifts. Errors include both the Poisson error associated with the observed number of galaxies and the errors in the estimated fraction of incorrect redshift measurements.

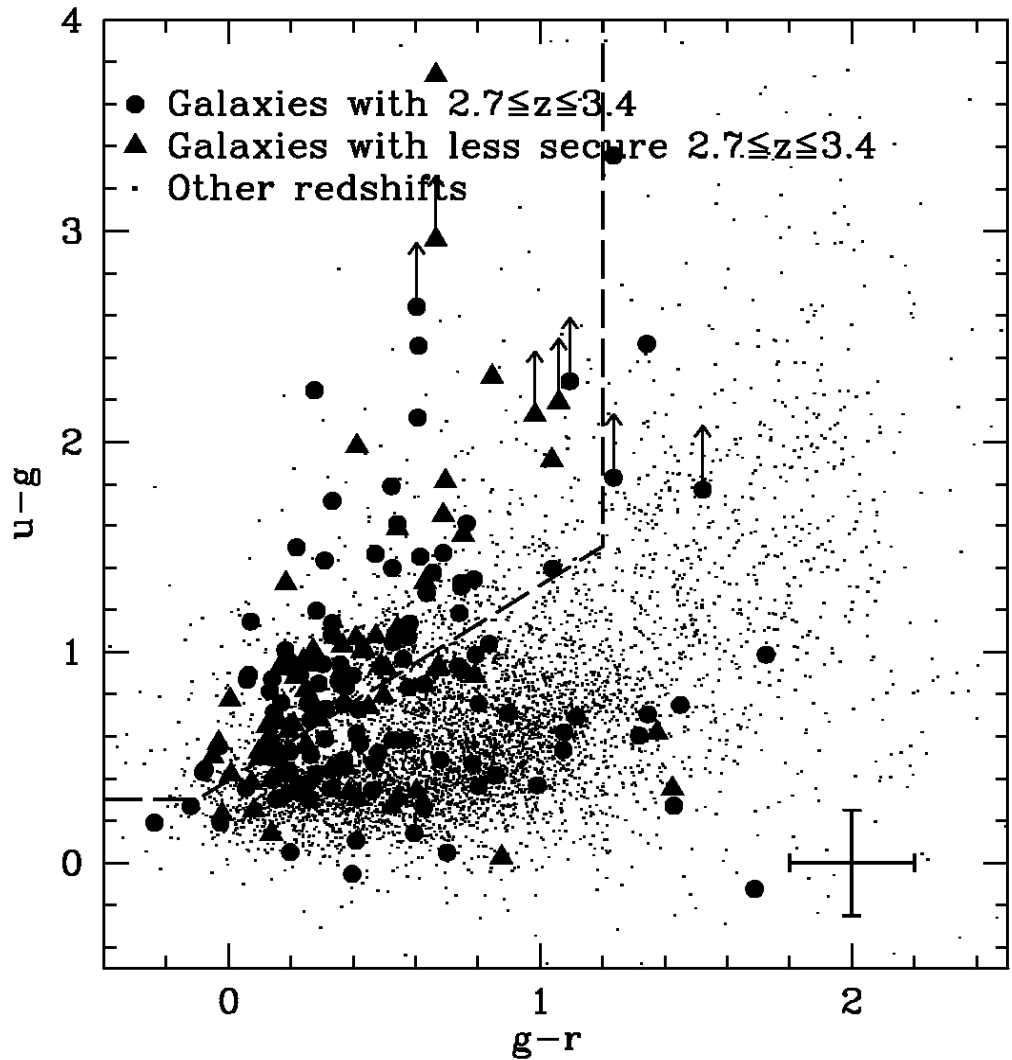

**Figure 3.** (*u−g*,*g−r*) **colour diagram of the VVDS high-redshift galaxies**. The *u-g* and *g-r* colours of galaxies are computed from their magnitudes in the u, g, and r bands (in the ultraviolet, green and red light, respectively). Galaxies with 2.7≤$z$≤3.4 are represented by circle symbols, and galaxies with less reliable redshifts in the same redshift range are represented by triangles. Galaxies at other redshifts are represented by small dots. The dashed line delineates the area where LBGs with 2.7≤$z$≤3.4 would be searched for if a colour–colour pre-selection was performed. The typical standard deviation errors on colours for the faintest galaxies are shown at the bottom right.